\documentclass{article}
\usepackage[T1]{fontenc} 
\usepackage[utf8]{inputenc} 
\usepackage{ismir,amsmath,cite,url}
\usepackage{graphicx}
\usepackage{color}

\usepackage{amsfonts}
\usepackage{enumitem}
\usepackage{booktabs}
\graphicspath{{figs/}}

\newcommand{\gcheck}{\checkmark}
\newcommand{\rcross}{}
\newcommand{\ytriangle}{{\footnotesize$\triangle$}}
\newcommand{\ccell}[1]{\multicolumn{1}{c}{#1}}

\title{MusPy: A Toolkit for Symbolic Music Generation}
\oneauthor{Hao-Wen Dong \hspace{1em} Ke Chen \hspace{1em} Julian McAuley \hspace{1em} Taylor Berg-Kirkpatrick}{University of California San Diego\\\tt\{hwdong, knutchen, jmcauley, tberg\}@ucsd.edu}
\def\authorname{Hao-Wen Dong, Ke Chen, Julian McAuley, Taylor Berg-Kirkpatrick}

\begin{document}

\maketitle

\begin{abstract}
In this paper, we present MusPy, an open source Python library for symbolic music generation. MusPy provides easy-to-use tools for essential components in a music generation system, including dataset management, data I/O, data preprocessing and model evaluation. In order to showcase its potential, we present statistical analysis of the eleven datasets currently supported by MusPy. Moreover, we conduct a cross-dataset generalizability experiment by training an autoregressive model on each dataset and measuring held-out likelihood on the others---a process which is made easier by MusPy's dataset management system. The results provide a map of domain overlap between various commonly used datasets and show that some datasets contain more representative cross-genre samples than others. Along with the dataset analysis, these results might serve as a guide for choosing datasets in future research. Source code and documentation are available at \url{https://github.com/salu133445/muspy}.
\end{abstract}

\section{Introduction}
\label{sec:introduction}

\begin{figure}
  \centering
  \includegraphics[width=\linewidth]{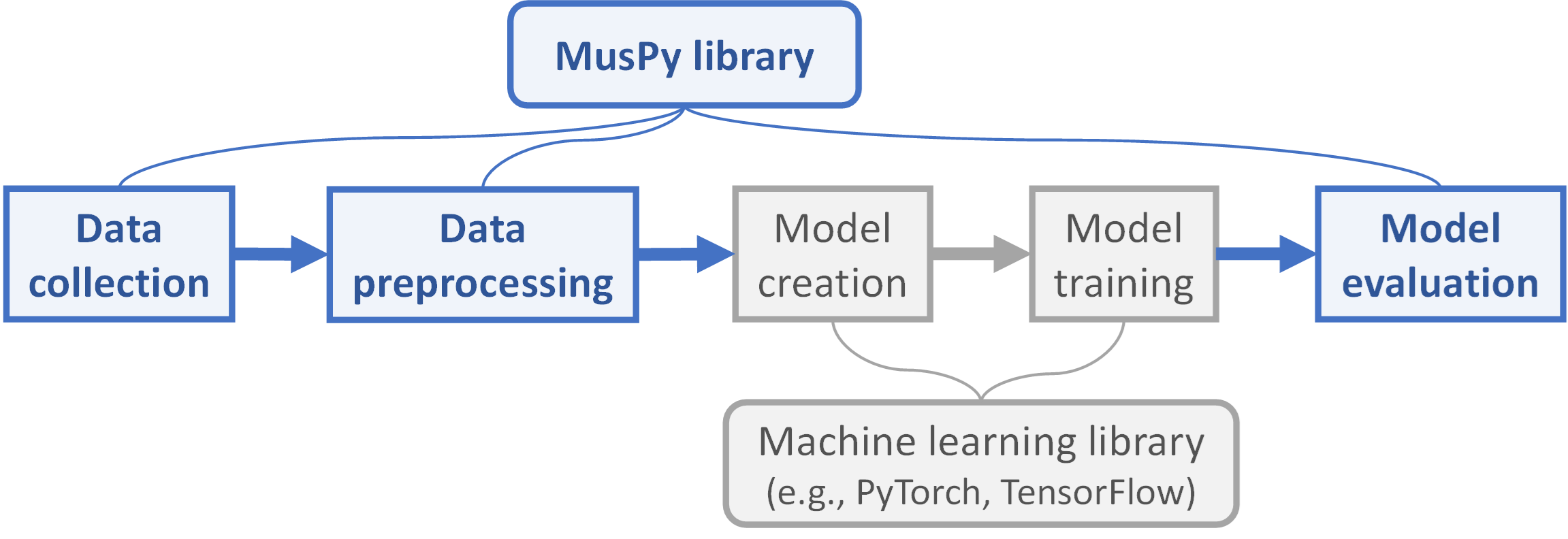}
  \caption{An example of a learning-based music generation system. MusPy provides basic routines specific to music as well as interfaces to machine learning frameworks.}
  \label{fig:pipeline}
\end{figure}

Recent years have seen progress on music generation, thanks largely to advances in machine learning~\cite{briot2017survey}. A music generation pipeline usually consists of several steps---data collection, data preprocessing, model creation, model training and model evaluation, as illustrated in \figref{fig:pipeline}. While some components need to be customized for each model, others can be shared across systems. For symbolic music generation in particular, a number of datasets, representations and metrics have been proposed in the literature~\cite{briot2017survey}. As a result, an easy-to-use toolkit that implements standard versions of such routines could save a great deal of time and effort and might lead to increased reproducibility. However, such tools are challenging to develop for a variety of reasons.

\begin{figure*}
  \centering
  \includegraphics[width=.98\linewidth]{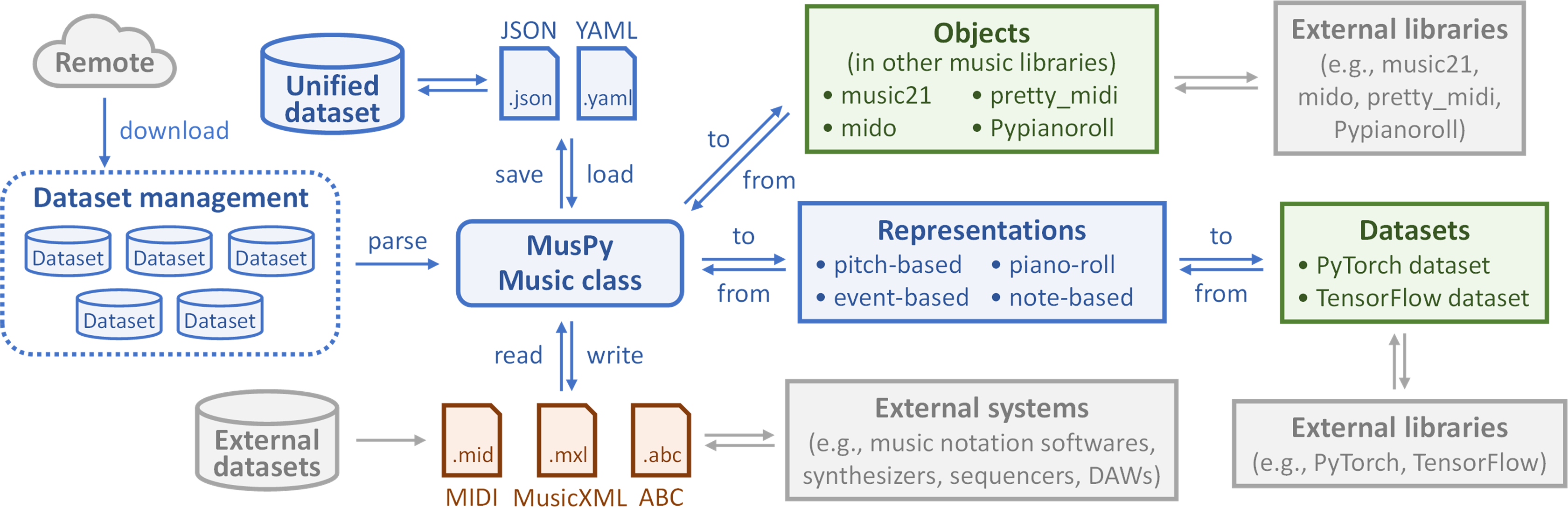}
  \caption{System diagram of MusPy. The MusPy Music object at the center is the core element of MusPy.}
  \label{fig:system}
\end{figure*}

First, though there are a number of publicly-available symbolic music datasets, the diverse organization of these collections and the various formats used to store them presents a challenge. 
These formats are usually designed for different purposes. Some focus on playback capability (e.g.,~MIDI), some are developed for music notation softwares (e.g.,~MusicXML~\cite{good2001musicxml} and LilyPond~\cite{lilypond}), some are designed for organizing musical documents (e.g.,~Music Encoding Initiative (MEI)~\cite{hankinson2011mei}), and others are research-oriented formats that aim for simplicity and readability (e.g.,~MuseData~\cite{hewlett1997musedata} and Humdrum~\cite{huron1997humdrum}. Oftentimes researchers have to implement their own preprocessing code for each different format. Moreover, while researchers can implement their own procedures to access and process the data, issues of reproducibility due to the inconsistency of source data have been raised in~\cite{bittner2019mirdata} for audio datasets.

Second, music has hierarchy and structure, and thus different levels of abstraction can lead to different representations~\cite{dannenberg1993survey}. Moreover, a number of music representations designed specially for generative modeling of music have also been proposed in prior art, for example, as a sequence of pitches~\cite{mozer1994neural,eck2002lstm,bl2012jsb,roberts2018musicvae}, events~\cite{oore2018rnn,huang2019transformer,donahue2019lakhnes,huang2020remi}, notes~\cite{mogren2016crnngan} or a time-pitch matrix (i.e.,~a piano roll)~\cite{yang2018midinet,dong2018musegan}.

Finally, efforts have been made toward more robust objective evaluation metrics for music generation systems~\cite{yang2018evaluation} as these metrics provide not only an objective way for comparing different models but also indicators for monitoring training progress in machine learning-based systems. Given the success of mir\_eval~\cite{raffel2014mireval} in evaluating common MIR tasks, a library providing implementations of commonly used evaluation metrics 
for music generation systems could help improve reproducibility.

To manage the above challenges, we find a toolkit dedicated for music generation a timely contribution to the MIR community. Hence, we present in this paper a new Python library, MusPy, for symbolic music generation. It provides essential tools for developing a music generation system, including dataset management, data I/O, data preprocessing and model evaluation.

With MusPy, we provide a statistical analysis on the eleven datasets currently supported by MusPy, with an eye to unveiling statistical differences between them. Moreover, we conduct three experiments to analyze their relative diversities and cross-dataset domain compatibility of the various datasets. These results, along with the statistical analysis, together provide a guide for choosing proper datasets for future research. Finally, we also show that combining multiple heterogeneous datasets could help improve generalizability of a music generation system.

\section{Related Work}
\label{sec:related-work}

Few attempts, to the best of our knowledge, have been made to develop a dedicated library for music generation. The Magenta project~\cite{magenta} represents the most notable example. While MusPy aims to provide fundamental routines in data collection, preprocessing and analysis, Magenta comes with a number of model instances, but is tightly bound with TensorFlow~\cite{abadi2016tensorflow}. In MusPy, we leave the model creation and training to dedicated machine learning libraries, and design MusPy to be flexible in working with different machine learning frameworks.

There are several libraries for working with symbolic music. music21~\cite{cuthbert2010music21} is one of the most representative toolkits and targets studies in computational musicology. While music21 comes with its own corpus, MusPy does not host any dataset. Instead, MusPy provides functions to download datasets from the web, along with tools for managing different collections, which makes it easy to extend support for new datasets in the future. jSymbolic~\cite{mckay2006jsymbolic} focuses on extracting statistical information from symbolic music data. While jSymbolic can serve as a powerful feature extractor for training supervised classification models, MusPy focuses on generative modeling of music and supports different commonly used representations in music generation. In addition, MusPy provides several objective metrics for evaluating music generation systems.

Related cross-dataset generalizability experiments~\cite{donahue2019lakhnes} show that pretraining on a cross-domain data can improve music generation results both qualitatively and quantitatively. MusPy's dataset management system makes it easier for us to thoroughly verify this hypothesis by examining pairwise generalizabilities between various datasets.

\section{MusPy}
\label{sec:muspy}

\begin{table*}
  \centering
  \begin{tabular}{llrrcccc}
    \toprule
    Dataset                                            &Format   &\ccell{Hours}  &\ccell{Songs}   &Genre     &Melody     &Chords     &Multitrack\\
    \midrule
    Lakh MIDI Dataset (LMD)~\cite{raffel16lmd}         &MIDI     &>9000  &174,533 &misc      &\ytriangle &\ytriangle &\ytriangle\\
    MAESTRO Dataset~\cite{hawthorne2018maestro}        &MIDI     &201.21 &1,282   &classical &\rcross    &\rcross    &\rcross\\
    Wikifonia Lead Sheet Dataset~\cite{wikifonia}      &MusicXML &198.40 &6,405   &misc      &\gcheck    &\gcheck    &\rcross\\
    Essen Folk Song Database~\cite{essen}              &ABC      &56.62  &9,034   &folk      &\gcheck    &\gcheck    &\rcross\\
    NES Music Database\cite{donahue2018nesmdb}         &MIDI     &46.11  &5,278   &game      &\gcheck    &\rcross    &\gcheck\\
    Hymnal Tune Dataset~\cite{hymnal}                  &MIDI     &18.74  &1,756  &hymn      &\gcheck    &\rcross    &\rcross\\
    Hymnal Dataset~\cite{hymnal}                       &MIDI     &17.50  &1,723   &hymn      &\rcross    &\rcross    &\rcross\\
    music21 Corpus\cite{cuthbert2010music21}         &misc     &16.86  &613     &misc      &\ytriangle &\rcross    &\ytriangle\\
    Nottingham Database (NMD)~\cite{nmd}               &ABC      &10.54  &1,036   &folk      &\gcheck    &\gcheck    &\rcross\\
    music21 JSBach Corpus~\cite{cuthbert2010music21} &MusicXML &3.46   &410     &classical &\rcross    &\rcross    &\gcheck\\
    JSBach Chorale Dataset~\cite{bl2012jsb}            &MIDI     &3.21   &382     &classical &\rcross    &\rcross    &\gcheck\\
    \bottomrule
  \end{tabular}
  \caption{Comparisons of datasets currently supported by MusPy. Triangle marks indicate partial support. Note that, in this version, only MusicXML and MIDI files are included for the music21 Corpus.}
  \label{tab:datasets}
\end{table*}

\begin{figure}
  \centering
  \begin{minipage}{.1\linewidth}
    (a)
  \end{minipage}
  \begin{minipage}{.85\linewidth}
    \includegraphics[width=\linewidth]{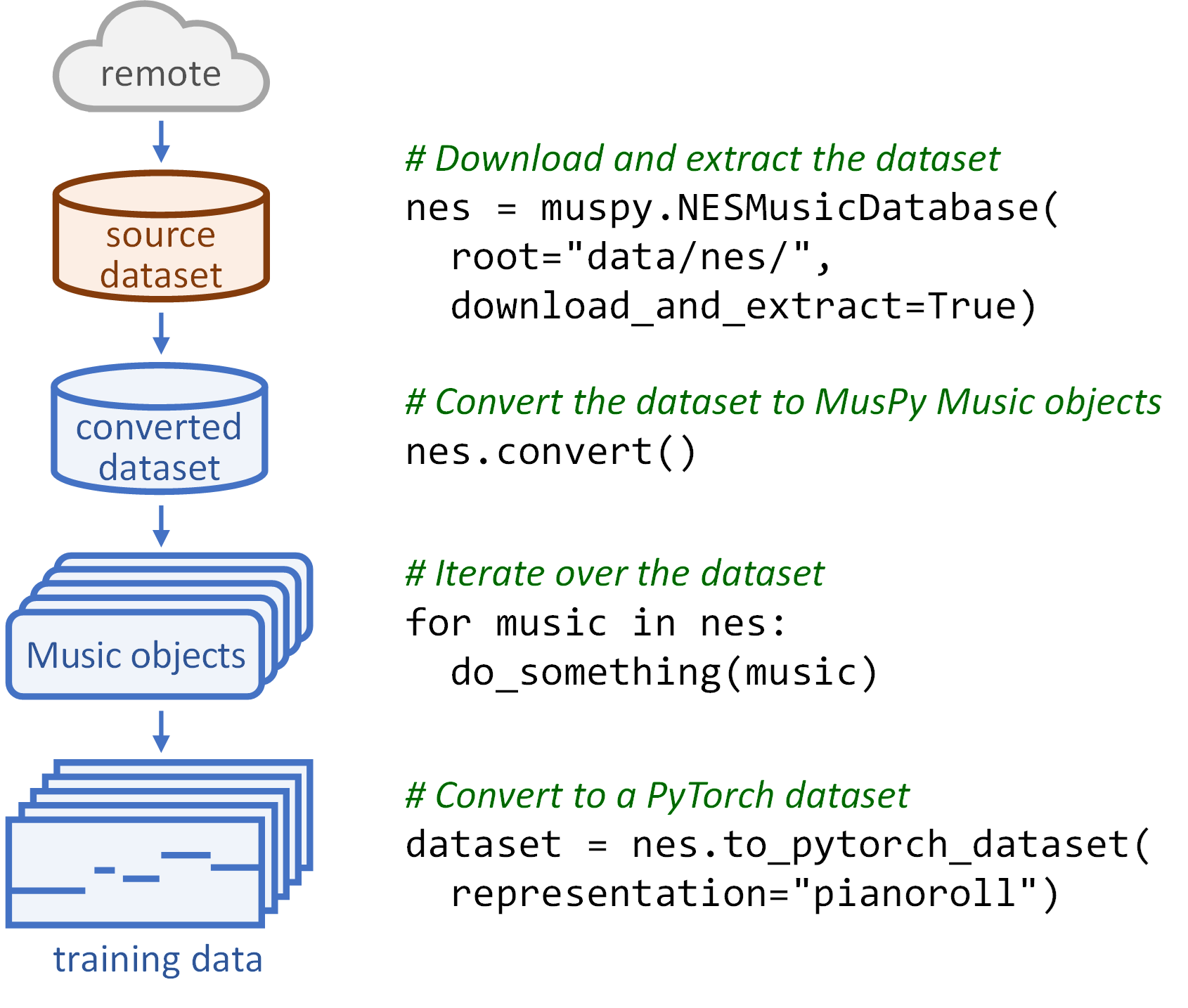}
  \end{minipage}\\[1ex]
  \begin{minipage}{.1\linewidth}
    (b)
  \end{minipage}
  \begin{minipage}{.85\linewidth}
    \includegraphics[width=\linewidth]{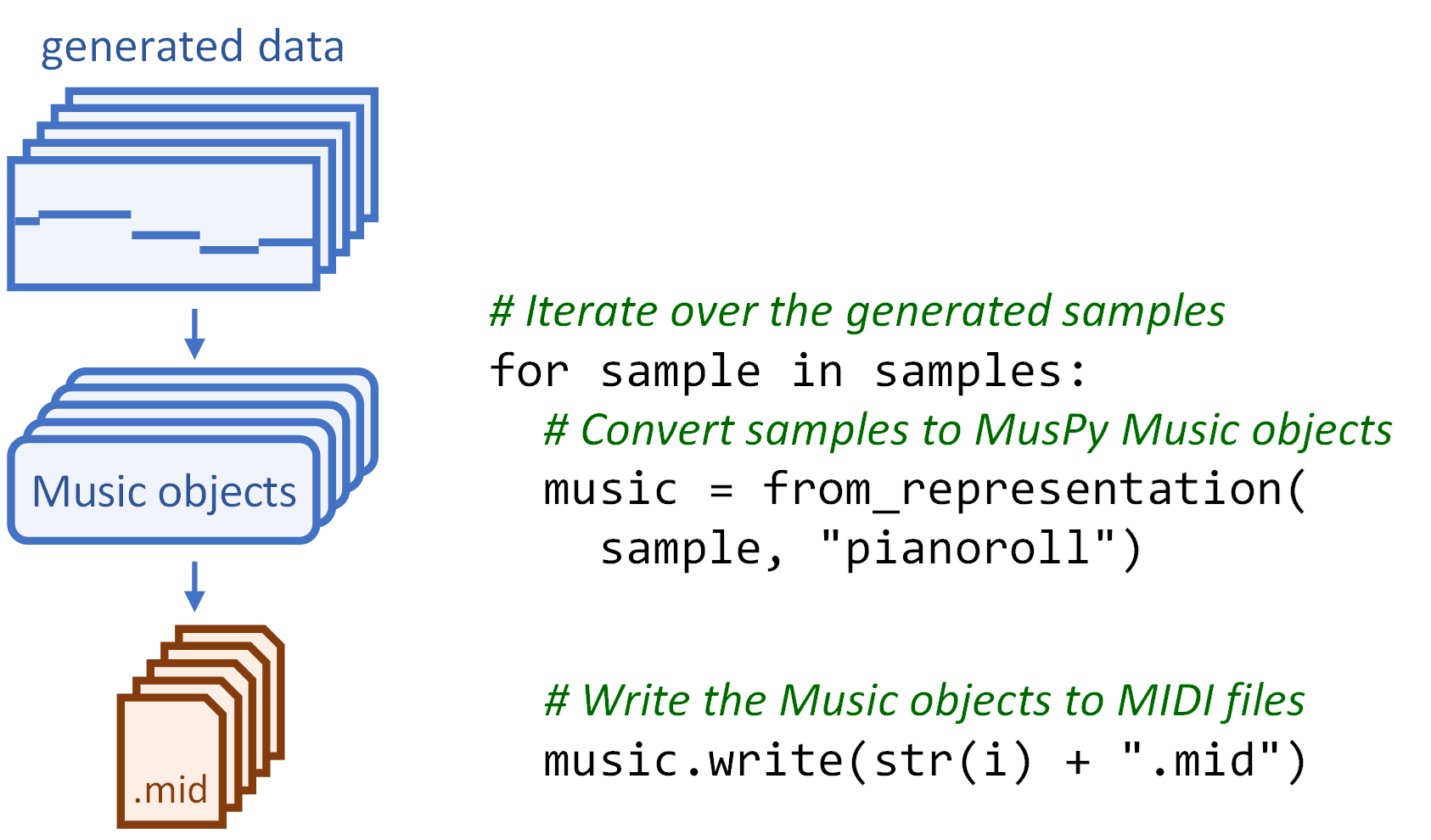}
  \end{minipage}
  \caption{Examples of (a) training data preparation and (b) result writing pipelines using MusPy.}
  \label{fig:pipeline_example}
\end{figure}

MusPy is an open source Python library dedicated for symbolic music generation. \figref{fig:system} presents the system diagram of MusPy. It provides a core class, MusPy Music class, as a universal container for symbolic music. Dataset management system, I/O interfaces and model evaluation tools are then built upon this core container. We provide in \figref{fig:pipeline_example} examples of data preparation and result writing pipelines using MusPy.

\subsection{MusPy Music class and I/O interfaces}
\label{sec:io}

We aim at finding a middle ground among existing formats for symbolic music and design a unified format dedicated for music generation. MIDI, as a communication protocol between musical devices, uses velocities to indicate dynamics, beats per minute (bpm) for tempo markings, and control messages for articulation, but it lacks the concepts of notes, measures and symbolic musical markings. In contrast, MusicXML, as a sheet music exchanging format, has the concepts of notes, measures and symbolic musical markings and contains visual layout information, but it falls short on playback-related data. For a music generation system, however, both symbolic and playback-specific data are important. Hence, we follow MIDI's standard for playback-related data and MusicXML's standard for symbolic musical markings.

\begin{table}
  \centering
  \begin{tabular}{l@{\hspace{3pt}}c@{\hspace{8pt}}c@{\hspace{8pt}}c}
    \toprule
                            &MIDI       &MusicXML   &MusPy\\
    \midrule
    Sequential timing       &\gcheck    &\rcross    &\gcheck\\
    Playback velocities     &\gcheck    &\ytriangle &\gcheck\\
    Program information     &\gcheck    &\ytriangle &\gcheck\\
    \midrule
    Layout information      &\rcross    &\gcheck    &\rcross\\
    Note beams and slurs    &\rcross    &\gcheck    &\rcross\\
    Song/source meta data   &\ytriangle &\gcheck    &\gcheck\\
    Track/part information  &\ytriangle &\gcheck    &\gcheck\\
    Dynamic/tempo markings  &\rcross    &\gcheck    &\gcheck\\
    Concept of notes        &\rcross    &\gcheck    &\gcheck\\
    Measure boundaries      &\rcross    &\gcheck    &\gcheck\\
    Human readability       &\rcross    &\ytriangle &\gcheck\\
    \bottomrule
  \end{tabular}
  \caption{Comparisons of MIDI, MusicXML and the proposed MusPy formats. Triangle marks indicate optional or limited support.}
  \label{tab:comparison}
\end{table}

In fact, the MusPy Music class naturally defines a universal format for symbolic music, which we will refer to as the MusPy format, and can be serialized into a human-readable JSON/YAML file. \tabref{tab:comparison} summarizes the key differences among MIDI, MusicXML and the proposed MusPy formats. Using the proposed MusPy Music class as the internal representation for music data, we then provide I/O interfaces for common formats (e.g., MIDI, MusicXML and ABC) and interfaces to other symbolic music libraries (e.g.,~music21~\cite{cuthbert2010music21}, mido~\cite{mido}, pretty\_midi~\cite{raffel2014prettymidi} and Pypianoroll~\cite{dong2018pypianoroll}). \figref{fig:pipeline_example}(b) provides an example of result writing pipeline using MusPy.

\subsection{Dataset management}
\label{sec:management}

\begin{figure}
  \centering
  \includegraphics[scale=0.19]{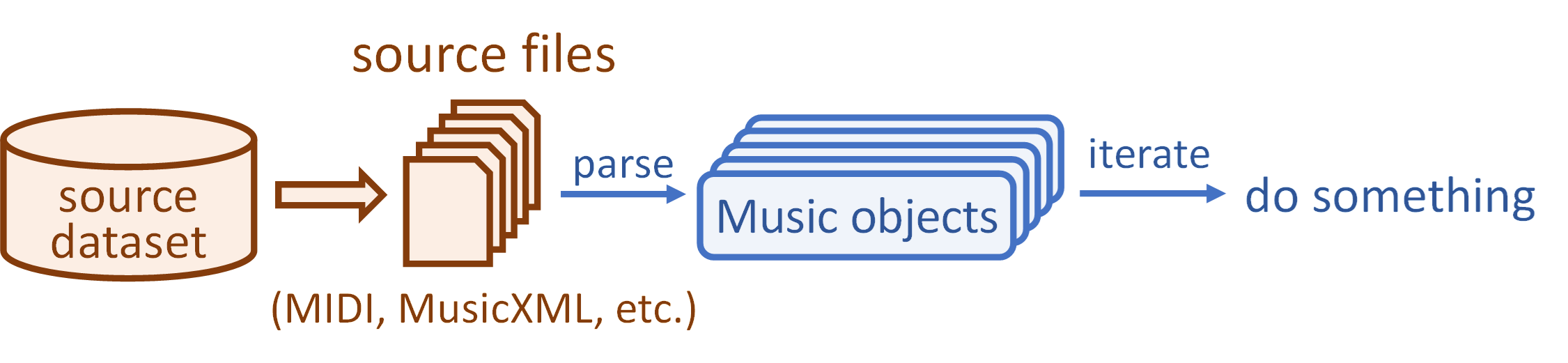}\\
  (a) on-the-fly mode\\[.5ex]
  \includegraphics[scale=0.19]{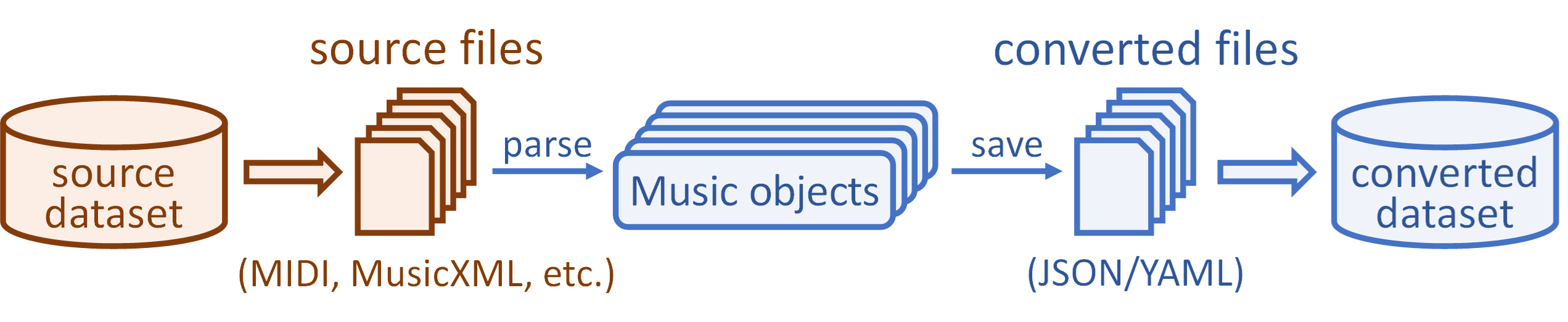}\\
  \includegraphics[scale=0.19]{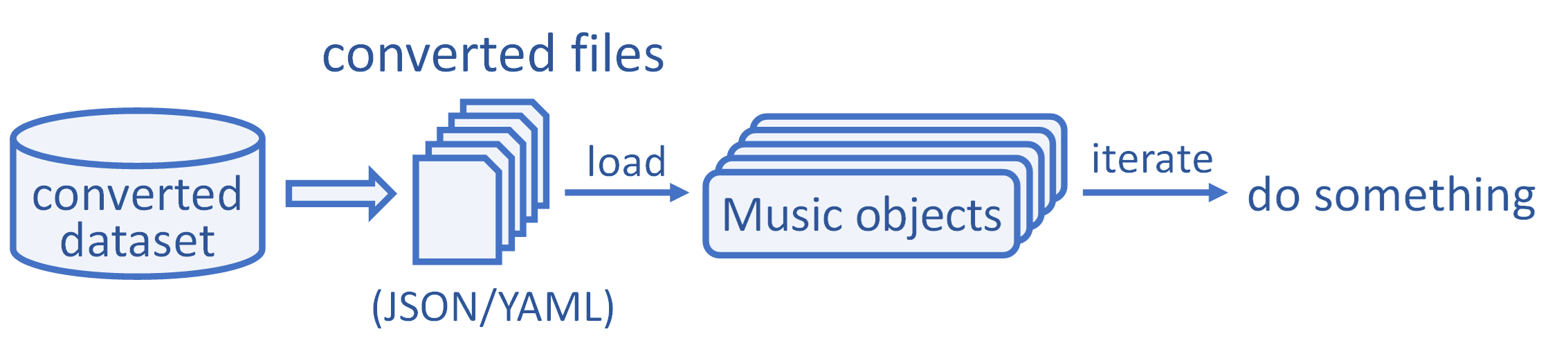}\\
  (b) preconverted mode
  \caption{Two internal processing modes for iterating over a MusPy Dataset object.}
  \label{fig:dataset_modes}
\end{figure}

MusPy provides an easy-to-use dataset management system similar to torchvision datasets~\cite{torchvision} and TensorFlow Dataset~\cite{tfds}. \tabref{tab:datasets} presents the list of datasets currently supported by MusPy and their comparisons. Each supported dataset comes with a class inherited from the base MusPy Dataset class. The modularized and flexible design of the dataset management system makes it easy to handle local data collections or extend support for new datasets in the future. \figref{fig:dataset_modes} illustrates the two internal processing modes when iterating over a MusPy Dataset object. In addition, MusPy provides interfaces to PyTorch~\cite{paszke2019pytorch} and TensorFlow~\cite{abadi2016tensorflow} for creating input pipelines for machine learning (see \figref{fig:pipeline_example}(a) for an example).

\begin{table*}
  \centering
  \begin{tabular}{llll}
    \toprule
    Representation &Shape &Values &Default configurations\\
    \midrule
    Pitch-based &$T \times 1$   &$\{0, 1,\dots, 129\}$          &$128$ note-ons, $1$ hold, $1$ rest (\textit{support only monophonic music})\\
    Event-based &$T \times 1$   &$\{0, 1,\dots, 387\}$          &$128$ note-ons, $128$ note-offs, $100$ time shifts, $32$ velocities\\
    Piano-roll   &$T \times 128$ &$\{0, 1\}$ or $\mathbb{R}^+$     &$\{0, 1\}$ for binary piano rolls; $\mathbb{R}^+$ for piano rolls with velocities\\
    Note-based  &$N \times 4$   &$\mathbb{N}$ or $\mathbb{R}^+$ &List of $(time, pitch, duration, velocity)$ tuples\\
    \bottomrule
  \end{tabular}
  \caption{Comparisons of representations supported by MusPy. $T$ and $N$ denote the numbers of time steps and notes, respectively. Note that the configurations can be modified to meet specific requirements and use cases.}
  \label{tab:representations}
\end{table*}

\subsection{Representations}

Music has multiple levels of abstraction, and thus can be expressed in various representations. For music generation in particular, several representations designed for generative modeling of symbolic music have been proposed and used in the literature~\cite{briot2017survey}. These representations can be broadly categorized into four types---the pitch-based~\cite{mozer1994neural,eck2002lstm,bl2012jsb,roberts2018musicvae}, the event-based~\cite{oore2018rnn,huang2019transformer,donahue2019lakhnes,huang2020remi}, the note-based~\cite{mogren2016crnngan} and the piano-roll~\cite{yang2018midinet,dong2018musegan} representations. \tabref{tab:representations} presents a comparison of them. We provide in MusPy implementations of these representations and integration to the dataset management system. \figref{fig:pipeline_example}(a) provides an example of preparing training data in the piano-roll representation from the NES Music Database using MusPy.

\subsection{Model evaluation tools}
\label{sec:rendering_visualization}

Model evaluation is another critical component in developing music generation systems. Hence, we also integrate into MusPy tools for audio rendering as well as score and piano-roll visualizations. These tools could also be useful for monitoring the training progress or demonstrating the final results. Moreover, MusPy provides implementations of several objective metrics proposed in the literature~\cite{mogren2016crnngan,dong2018musegan,wu2020jazz}. These objective metrics, as listed below, could be used to evaluate a music generation system by comparing the statistical difference between the training data and the generated samples, as discussed in~\cite{yang2018evaluation}. 
\begin{itemize}[leftmargin=*,itemsep=0pt]
  \item \textit{Pitch-related metrics}---polyphony, polyphony rate, pitch-in-scale rate, scale consistency, pitch entropy and pitch class entropy.
  \item \textit{Rhythm-related metrics}---empty-beat rate, drum-in-pattern rate, drum pattern consistency and groove consistency.
\end{itemize}

\subsection{Summary}
\label{sec:summary}

To summarize, MusPy features the following:
\begin{itemize}[leftmargin=*,itemsep=0pt]
  \item Dataset management system for commonly used datasets with interfaces to PyTorch and TensorFlow.
  \item Data I/O for common symbolic music formats (e.g., MIDI, MusicXML and ABC) and interfaces to other symbolic music libraries (e.g., music21, mido, pretty\_midi and Pypianoroll).
  \item Implementations of common music representations for music generation, including the pitch-based, the event-based, the piano-roll and the note-based representations.
  \item Model evaluation tools for music generation systems, including audio rendering, score and piano-roll visualizations and objective metrics.
\end{itemize}
All source code and documentation can be found at \url{https://github.com/salu133445/muspy}.

\section{Dataset Analysis}
\label{sec:analysis}

\begin{figure}
  \centering
  \includegraphics[width=\linewidth]{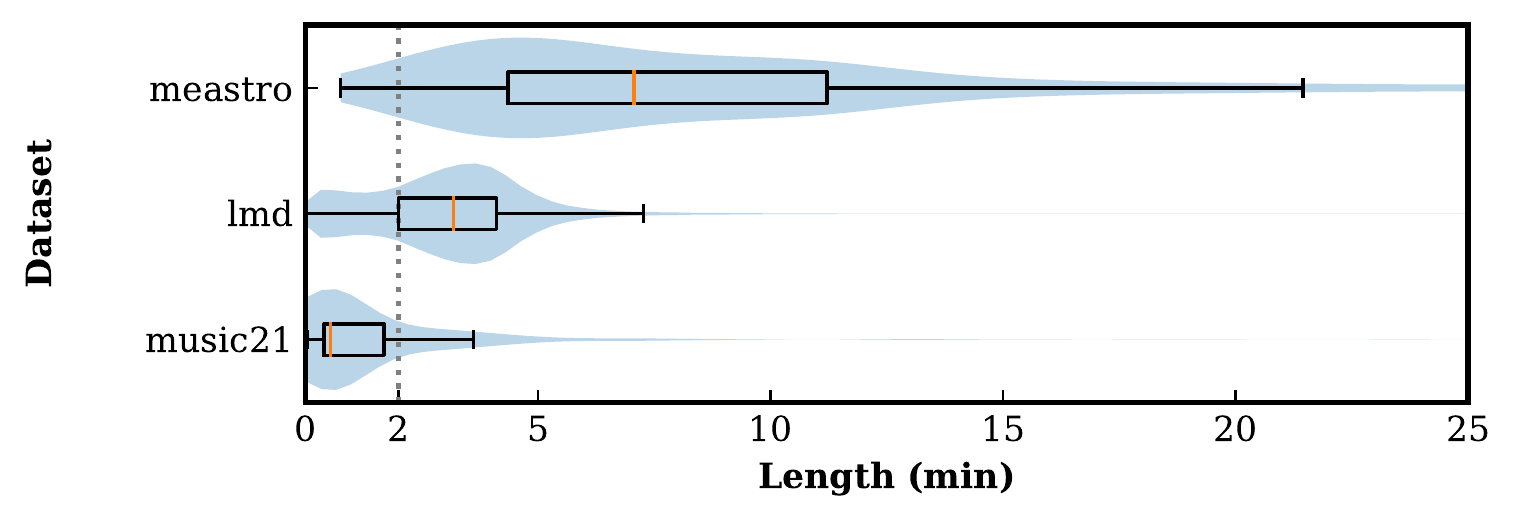}\\
  \includegraphics[width=\linewidth]{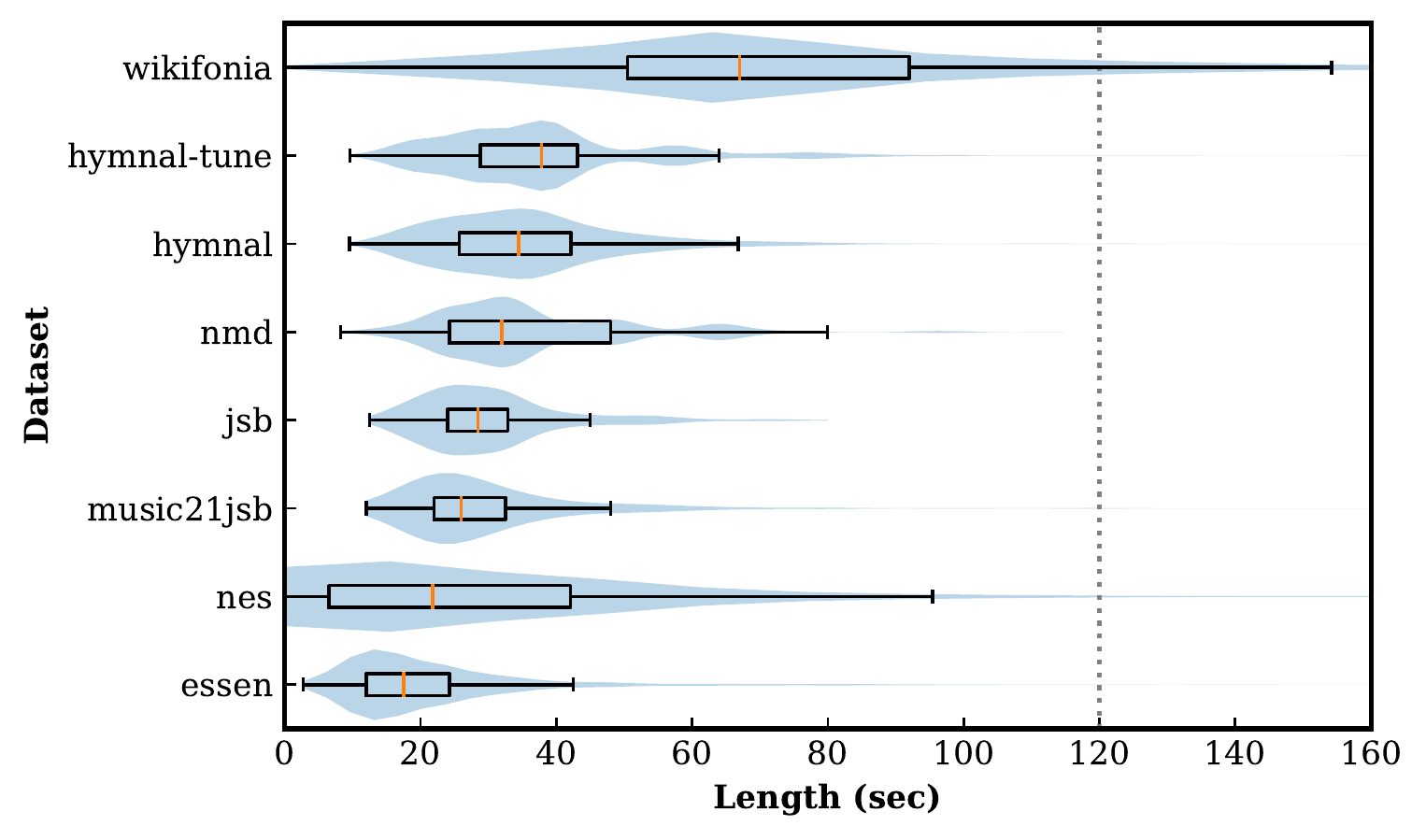}
  \caption{Length distributions for different datasets.}
  \label{fig:length_dist}
\end{figure}

Analyzing datasets is critical in developing music generation systems. With MusPy's dataset management system, we can easily work with different music datasets. Below we compute the statistics of three key elements of a song---length, tempo and key using MusPy, with an eye to unveiling statistical differences among these datasets. First, \figref{fig:length_dist} shows the distributions of song lengths for different datasets. We can see that they differ greatly in their ranges, medians and variances.

\begin{figure}
  \centering
  \includegraphics[width=\linewidth]{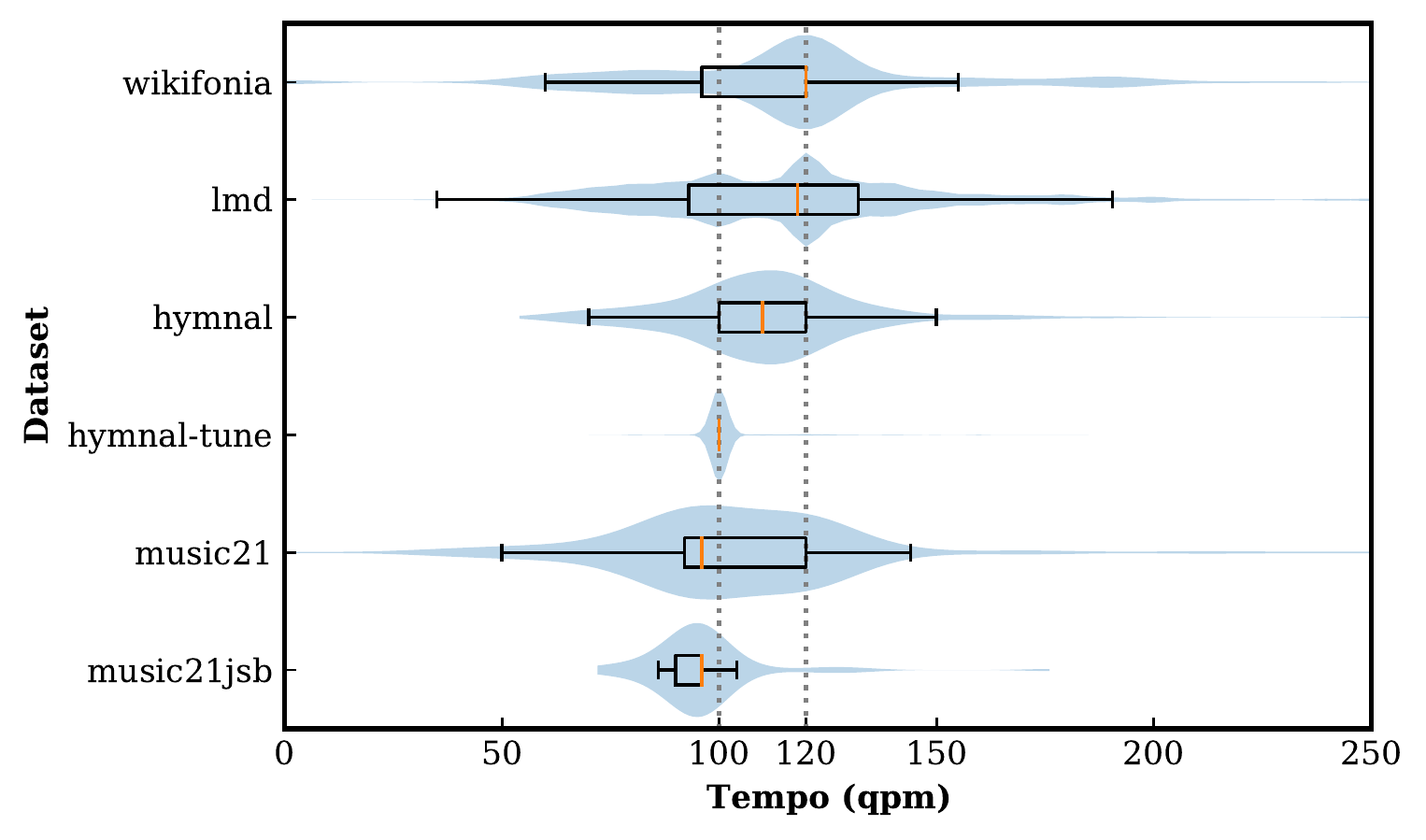}
  \caption{Initial-tempo distributions for different datasets (those without tempo information are not presented).}
  \label{fig:tempo_dist}
\end{figure}

Second, we present in \figref{fig:tempo_dist} the distributions of initial tempo for datasets that come with tempo information. We can see that all of them are generally bell-shaped but with different ranges and variances. We also note that there are two peaks, $100$ and $120$ quarter notes per minute (qpm), in Lakh MIDI Dataset (LMD), which is possibly because these two values are often set as the default tempo values in music notation programs and MIDI editors/sequencers. Moreover, in Hymnal Tune Dataset, only around ten percent of songs have an initial tempo other than $100$ qpm.

\begin{figure}
  \centering
  \includegraphics[width=\linewidth]{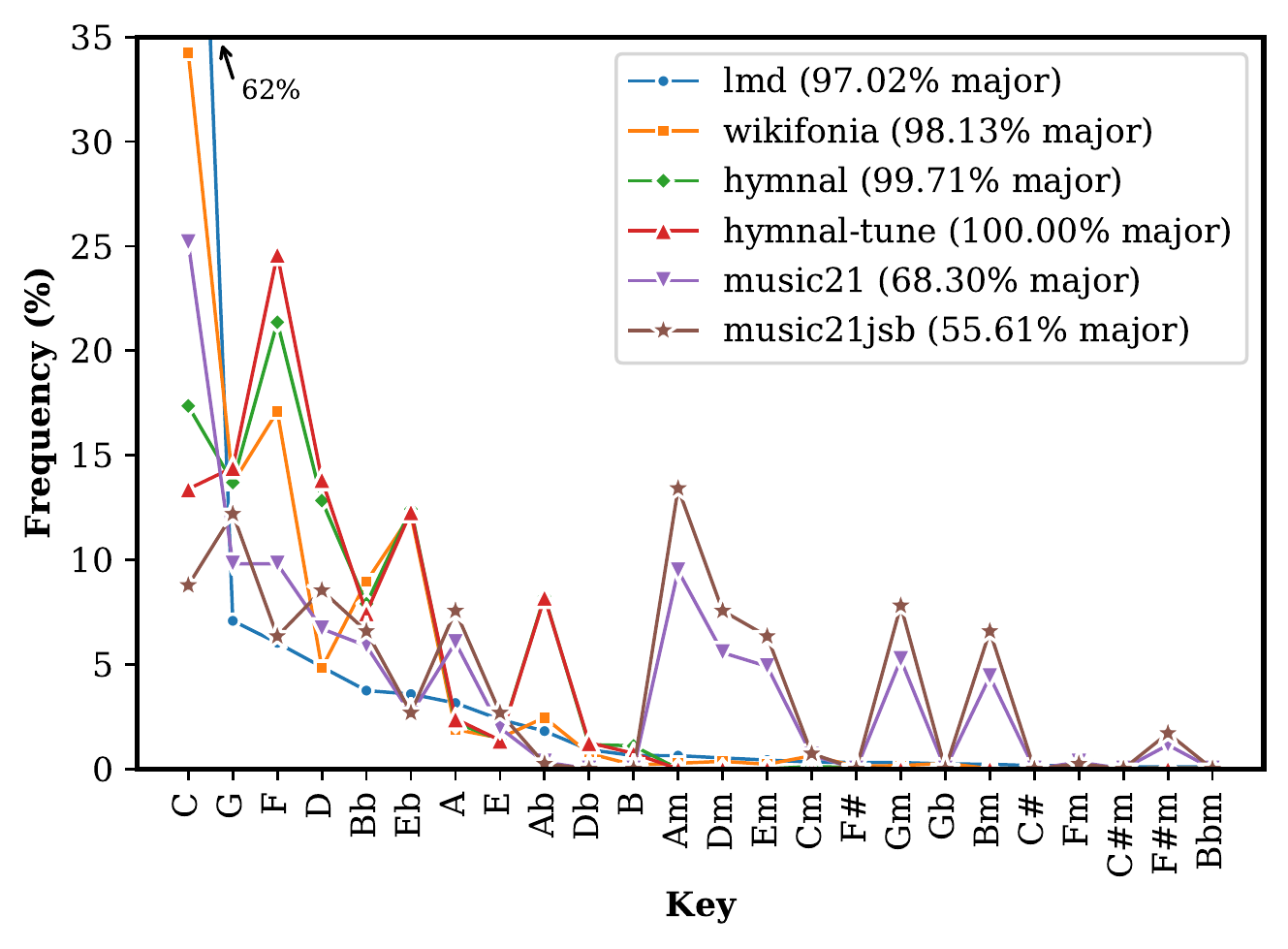}
  \caption{Key distributions for different datasets. The keys are sorted w.r.t.~their frequencies in Lakh MIDI Dataset.}
  \label{fig:key_hist}
\end{figure}

Finally, \figref{fig:key_hist} shows the histograms of keys for different datasets. We can see that the key distributions are rather imbalanced. Moreover, only less than 3\% of songs are in minor keys for most datasets except the music21 Corpus. In particular, LMD has the most imbalanced key distributions, which might be due to the fact that C major is often set as the default key in music notation programs and MIDI editors/sequencers.\footnote{Note that key information is considered as a meta message in a MIDI file. It does not affect the playback and thus can be unreliable sometimes.} These statistics could provide a guide for choosing proper datasets in future research.

\section{Experiments and Results}
\label{sec:exp}

In this section, we conduct three experiments to analyze the relative complexities and the cross-dataset generalizabilities of the eleven datasets currently supported by MusPy (see \tabref{tab:datasets}). We implement four autoregressive models---a recurrent neural network (RNN), a long short-term memory (LSTM) network~\cite{hochreiter1997lstm}, a gated recurrent unit (GRU) network~\cite{cho2014gru} and a Transformer network~\cite{vaswani2017transformer}.

\subsection{Experiment settings}
\label{sec:exp_settings}

For the data, we use the event representation as specified in \tabref{tab:representations} and discard velocity events as some datasets have no velocity information (e.g.,~datasets using ABC format). Moreover, we also include an end-of-sequence event, leading to in total $357$ possible events. For simplicity, we downsample each song into four time steps per quarter note and fix the sequence length to $64$, which is equivalent to four measures in $4/4$ time. In addition, we discard repeat information in MusicXML data and use only melodies in Wikifonia dataset. We split each dataset into train--test--validation sets with a ratio of $8:1:1$. For the training, the models are trained to predict the next event given the previous events. We use the cross entropy loss and the Adam optimizer~\cite{kingma2015adam}. For evaluation, we randomly sample $1000$ sequences of length $64$ from the test split, and compute the perplexity of these sequences. We implement the models in Python using PyTorch. For reproducibility, source code and hyperparmeters are available at {\small\url{https://github.com/salu133445/muspy-exp}}.

\subsection{Autoregressive models on different datasets}
\label{sec:exp_datasets}

\begin{figure}
  \centering
  \includegraphics[width=\linewidth]{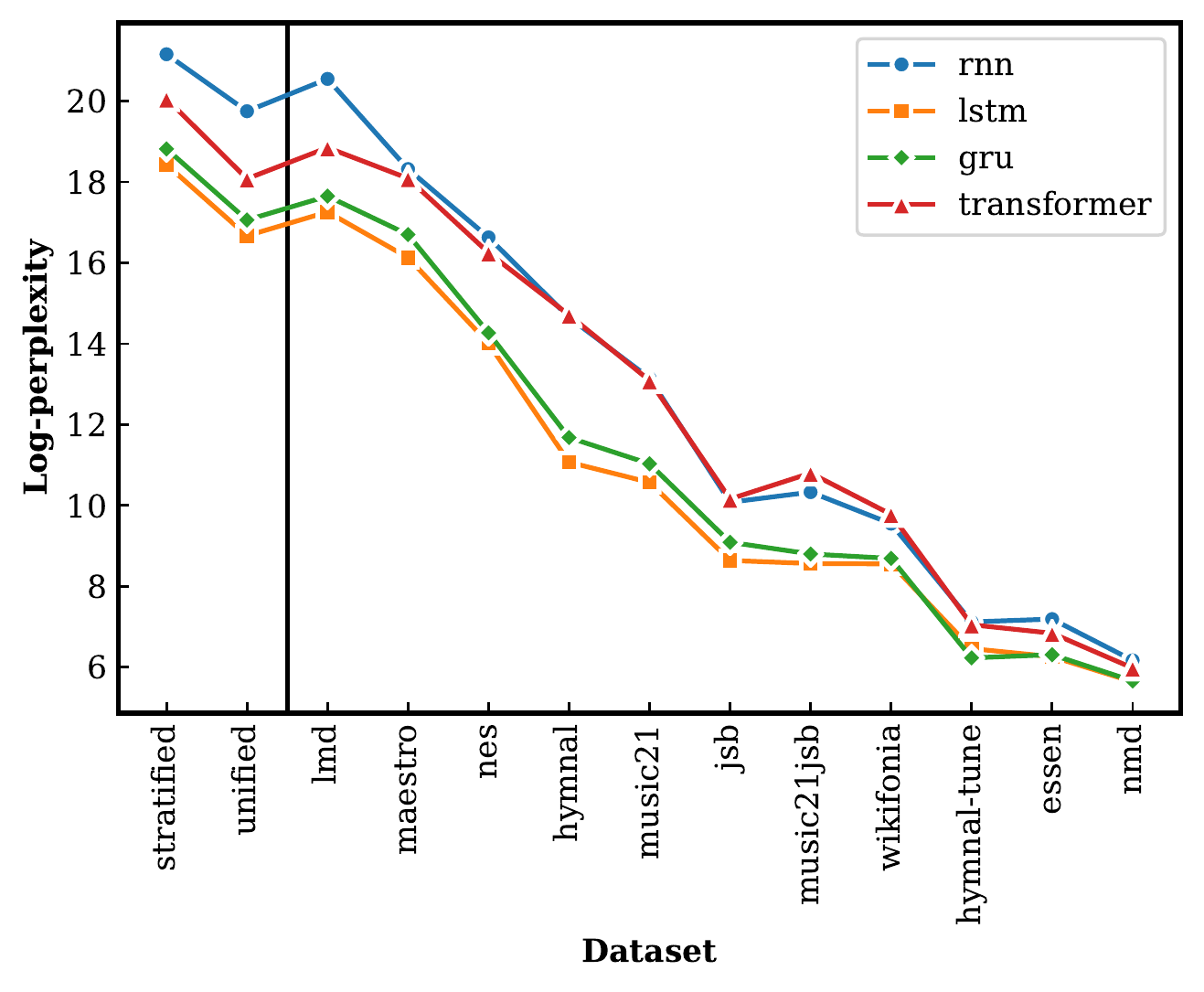}
  \caption{Log-perplexities for different models on different datasets, sorted by the values for the LSTM model.}
  \label{fig:exp_perplexity}
\end{figure}

\begin{figure}
  \centering
  \includegraphics[width=\linewidth]{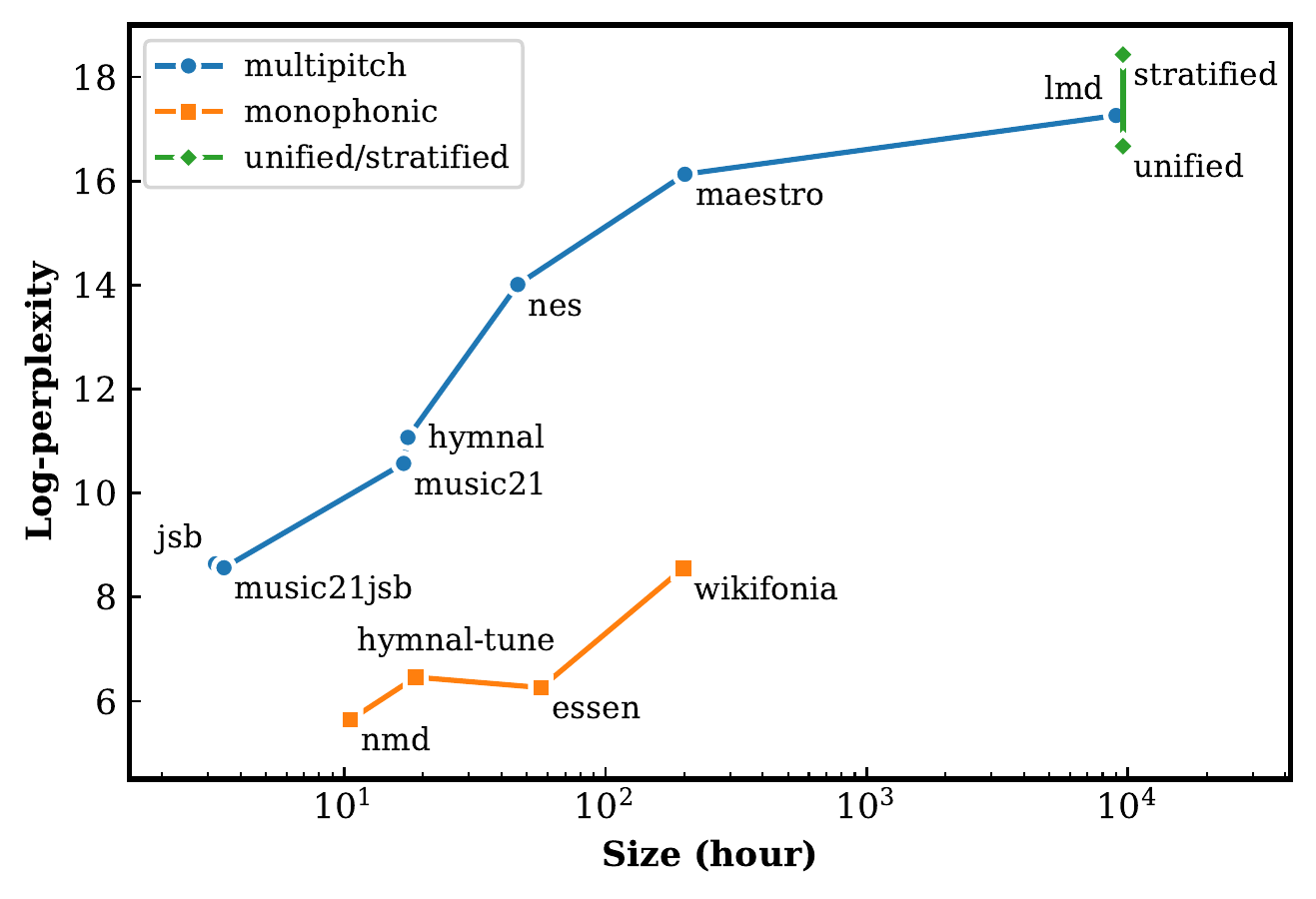}
  \caption{Log-perplexities for the LSTM model versus dataset size in hours. Each point corresponds to a dataset.}
  \label{fig:exp_hour_perplexity}
\end{figure}

In this experiment, we train the model on some dataset $\mathcal{D}$ and test it on the same dataset $\mathcal{D}$. We present in \figref{fig:exp_perplexity} the perplexities for different models on different datasets. We can see that all models have similar tendencies. In general, they achieve smaller perplexities for smaller, homogeneous datasets, but result in larger perplexities for larger, more diverse datasets. That is, the test perplexity could serve as an indicator for the diversity of a dataset. Moreover, \figref{fig:exp_hour_perplexity} shows perplexities versus dataset sizes (in hours). By categorizing datasets into multi-pitch (i.e.,~accepting any number of concurrent notes) and monophonic datasets, we can see that the perplexity is positively correlated to the dataset size within each group.

\subsection{Cross-dataset generalizability}
\label{sec:exp_cross_datasets}

In this experiment, we train a model on some dataset $\mathcal{D}$, while in addition to testing it on the same dataset $\mathcal{D}$, we also test it on each other dataset $\mathcal{D'}$. We present in \figref{fig:exp_cross_datasets} the perplexities for each train--test dataset pair. Here are some observations:
\begin{itemize}[leftmargin=*,itemsep=0pt]
  \item Cross dataset generalizability is not symmetric in general. For example, a model trained on LMD generalizes well to all other datasets, while not all models trained on other datasets generalize to LMD, which is possibly due to the fact that LMD is a large, cross-genre dataset.
  \item Models trained on multi-pitch datasets generalize well to monophonic datasets, while models trained on monophonic datasets do not generalize to multi-pitch datasets (see the red block in~\figref{fig:exp_cross_datasets}).
  \item The model trained on JSBach Chorale Dataset does not generalize to any of the other datasets (see the orange block in~\figref{fig:exp_cross_datasets}). This is possibly because its samples are downsampled to a resolution of quarter note, which leads to a distinct note duration distribution.
  \item Most datasets generalize worse to NES Music Database compared to other datasets (see the green block in~\figref{fig:exp_cross_datasets}). This is possibly due to the fact that NES Music Database contains only game soundtracks.
\end{itemize}

\subsection{Effects of combining heterogeneous datasets}
\label{sec:exp_large_dataset}

\begin{figure}
  \centering
  \includegraphics[width=\linewidth]{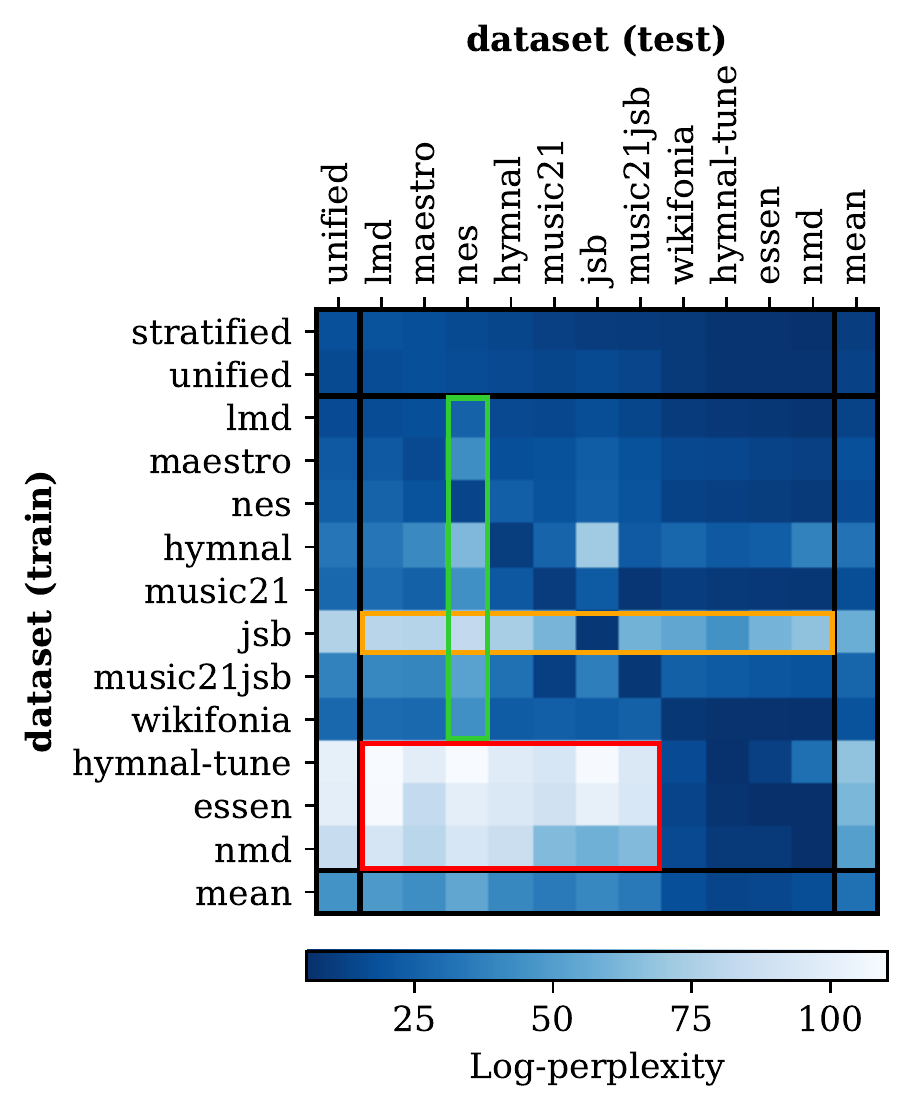}\\
  \includegraphics[width=\linewidth]{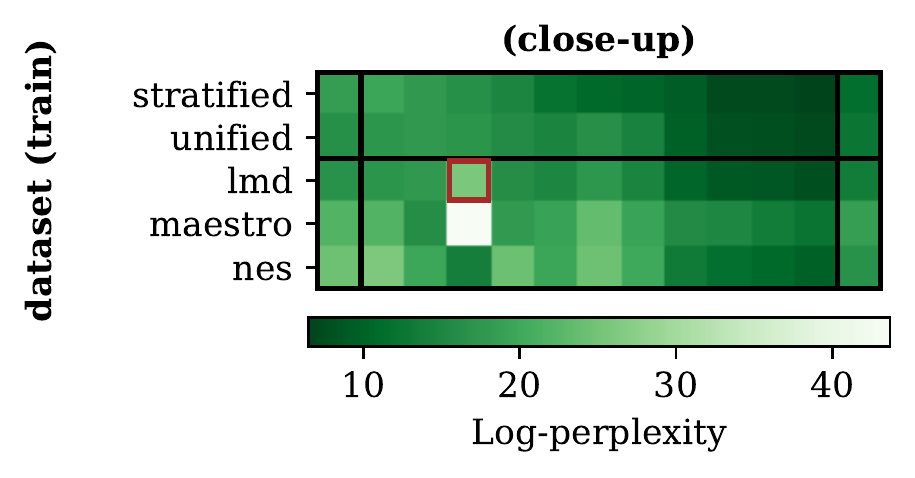}
  \caption{Cross-dataset generalizability results. The values and colors represent the log-perplexities of a LSTM model trained on a specific dataset (row) and tested on another dataset (column). The datasets are sorted by the diagonal values, i.e., trained and tested on the same dataset.}
  \label{fig:exp_cross_datasets}
\end{figure}

From \figref{fig:exp_cross_datasets} we can see that LMD has the best generalizability, possibly because it is large, diverse and cross-genre. However, a model trained on LMD does not generalize well to NES Music Database (see the brown block in the close-up of \figref{fig:exp_cross_datasets}). We are thus interested in whether combing multiple heterogeneous datasets could help improve generalizability.

We combine all eleven datasets listed in \tabref{tab:datasets} into one large \textit{unified} dataset. Since these datasets differ greatly in their sizes, simply concatenating the datasets might lead to severe imbalance problem and bias toward the largest dataset. Hence, we also consider a version that adopts stratified sampling during training. Specifically, to acquire a data sample in the \textit{stratified} dataset, we uniformly choose one dataset out of the eleven datasets, and then randomly pick one sample from that dataset. Note that stratified sampling is disabled at test time.

We also include in Figures \ref{fig:exp_perplexity}, \ref{fig:exp_hour_perplexity} and \ref{fig:exp_cross_datasets} the results for these two datasets. We can see from \figref{fig:exp_cross_datasets} that combining datasets from different sources improves the generalizability of the model. This is consistent with the finding in \cite{donahue2019lakhnes} that models trained on certain cross-domain datasets generalize better to other unseen datasets. Moreover, stratified sampling alleviates the source imbalance problem by reducing perplexities in most datasets with a sacrifice of an increased perplexity on LMD.

\section{Conclusion}
\label{sec:conclusion}

We have presented MusPy, a new toolkit that provides essential tools for developing music generation systems. We discussed the designs and features of the library, along with data pipeline examples. With MusPy's dataset management system, we conducted a statistical analysis and experiments on the eleven currently supported datasets to analyze their relative diversities and cross-dataset generalizabilities. These results could help researchers choose appropriate datasets in future research. Finally, we showed that combining heterogeneous datasets could help improve generalizability of a machine learning model.

\bibliography{ref}

\end{document}